\documentclass[amsmath,amssymb,superscriptaddress,nobalancelastpage,prb,twocolumn]{revtex4-2}
\usepackage{hyperref}
\usepackage{graphicx}
\usepackage{varioref}
\usepackage{xr-hyper}
\usepackage{xcolor}
\usepackage{nicefrac}
\usepackage{xfrac}
\usepackage{hyperref}
\hypersetup{colorlinks,linkcolor=blue,urlcolor=blue,citecolor=blue}
\usepackage{ulem}
\usepackage{lineno}
\usepackage{amsmath}
\usepackage{amssymb}
\usepackage{siunitx}

\def \CCA{CeCu$_{5.8}$Ag$_{0.2}$}

\makeatletter
\def\maketitle{
\@author@finish
\title@column\titleblock@produce
\suppressfloats[t]}
\makeatother

\graphicspath{{Images/}} % Path to the image folder

\begin{document}

\title{Magnetic Field Dependence of the Spin Fluctuations in CeCu$_{5.8}$Ag$_{0.2}$}
\date{\today}

\author{X. Boraley}
\affiliation{PSI Center for Neutron and Muon Sciences, 5232 Villigen PSI, Switzerland}

\author{A. D. Christianson}
\email{christiansad@ornl.gov}
\affiliation{Materials Science and Technology Division, Oak Ridge National Laboratory, Oak Ridge, Tennessee 37831, USA}

\author{J. Lass}
\affiliation{PSI Center for Neutron and Muon Sciences, 5232 Villigen PSI, Switzerland}

\author{C. Balz}
\affiliation{ISIS Neutron and Muon Source, STFC Rutherford Appleton Laboratory, Didcot OX11 0QX, United Kingdom}
\affiliation{Neutron Scattering Division, Oak Ridge National Laboratory, Oak Ridge, Tennessee 37831, USA}

\author{M. Bartkowiak}
\affiliation{PSI Center for Neutron and Muon Sciences, 5232 Villigen PSI, Switzerland}

\author{Ch. Niedermayer}
\affiliation{PSI Center for Neutron and Muon Sciences, 5232 Villigen PSI, Switzerland}

\author{J. M. Lawrence}
\affiliation{Los Alamos National Laboratory, Los Alamos, NM, 87545, USA}

\author{L. Poudel}
\affiliation{Neutron Scattering Division, Oak Ridge National Laboratory, Oak Ridge, Tennessee 37831, USA}
\affiliation{Department of Physics \& Astronomy, University of Tennessee, Knoxville, TN 37996, USA}

\author{ D. G. Mandrus}
\affiliation{Department of Physics \& Astronomy, University of Tennessee, Knoxville, TN 37996, USA}
\affiliation{Materials Science \& Technology Division, Oak Ridge National Laboratory, Oak Ridge, TN 37831, USA}
\affiliation{Department of Materials Science \& Engineering, University of Tennessee, Knoxville, TN 37996, USA}

\author{F. Ronning}
\affiliation{Los Alamos National Laboratory, Los Alamos, NM, 87545, USA}

\author{M. Janoschek}
\affiliation{PSI Center for Neutron and Muon Sciences, 5232 Villigen PSI, Switzerland}
\affiliation{Physik-Institut, Universit{\"a}t Zurich, Winterthurerstrasse 190, CH-8057 Zurich, Switzerland}

\author{D. G. Mazzone}
\email{daniel.mazzone@psi.ch}
\affiliation{PSI Center for Neutron and Muon Sciences, 5232 Villigen PSI, Switzerland}

\begin{abstract}
Quantum phase transitions are among the most intriguing phenomena that can occur when the electronic ground state of correlated metals are tuned by external parameters such as pressure, magnetic field or chemical substitution. Such transitions between distinct states of matter are driven by quantum fluctuations, and can give rise to macroscopically coherent phases that are at the forefront of condensed matter research. However, the nature of the critical fluctuations, and thus the fundamental physics controlling  many quantum phase transitions, remain poorly understood in numerous strongly correlated metals. Here we study the model material CeCu$_{5.8}$Ag$_{0.2}$ to gain insight into the implications of critical fluctuations originating from different regions in reciprocal space. By employing an external magnetic field along the crystallographic $a$- and $c$-axis as auxiliary tuning parameter we observe a pronounced anisotropy in the suppression of the quantum critical fluctuations,  reflecting the spin anisotropy of the
long-range ordered ground state at larger silver concentration. Coupled with the temperature dependence of the quantum fluctuations, these results suggest that the quantum phase transition in CeCu$_{5.8}$Ag$_{0.2}$ is driven by three-dimensional spin-density wave fluctuations.

\end{abstract}

\maketitle

%----------------------------------------------------------------------
\section{Introduction}

A thermodynamic system of interacting particles assumes a variety of states of matter ranging from different configurations in solids to liquids, gases and plasmas. Transformations between configurations occur as external conditions change, such as a solid that melts into a liquid at elevated temperature. Here we focus on continuous processes or phase transitions that can be driven either by thermal or quantum fluctuations. The latter fluctuations are particularly interesting, because they can trigger novel coherent states in solids that are potentially appealing for future technological applications \cite{Vojta_2018,Savary2017,Norman2011,Narayan2019}. Quantum phase transitions (QPTs) are governed by Heisenberg's uncertainty principle, conceptually distinguishing them from classical transitions.  
Quantum  critical fluctuations diverge in space and imaginary time \cite{Wilson1974, Sachdev_2011}, elevating the effective dimension of the dynamic properties around the transition. These higher-dimensional fluctuations substantially alter the material properties around the QPT as has been shown by macroscopic resistivity, heat capacity, susceptibility, Gr\"uneisen measurements or through inelastic neutron scattering \cite{Tokiwa2015, Vivek2018,Singh_2011,Tran2012,Knafo_2009}. Their comparison to theoretical models allows for a detailed assessment of the quantum critical nature of the fluctuations \cite{Hertz_1976,Millis_1993,Moriya_1995,Si_2001,Si_2013,Sachdev1999,Kirkpatrick2015,Vojta_2018}. However, in many metals the experimental observations disagree with the prevailing theoretical predictions of quantum fluctuations \cite{Schroder_1998,Schroder_2000,Lohneysen_2007,Park2002,Aronson1995,Stock2012}.

A central challenge in metals is that multiple degrees of freedom such as spin, charge and lattice may become quantum critical. Long-range magnetic order is predominantly mediated via the RKKY interaction \cite{Ruderman_1954,Kasuya_1956,Yoshida_1957}, in which the conduction electrons act as the exchange pathway between the localized ions. The prevailing theory derived by Hertz, Millis and Moriya (HMM) predicts five dimensional quantum critical fluctuations for such a spin-density wave (SDW) quantum critical point \cite{Hertz_1976,Millis_1993,Moriya_1995}, yielding mean-field behavior even close to the phase transition. Experimental realizations of SDW QPTs have been found in Ce$_{1-x}$La$_{x}$Ru$_{2}$Si$_{2}$ and in CeCu$_{2}$Ge$_{2}$ under magnetic field, for instance \cite{Knafo_2009, Singh_2011}. However, in materials containing Ce, Yb, Sm, U and other rare earths the conduction electrons can also screen the local moments via the Kondo effect \cite{Kondo_1964}. A QPT associated with a quantum critical Kondo breakdown leads to local quantum critical fluctuations \cite{Si_2001, Si_2013}, which are thought to be at play in CeRhIn$_5$ \cite{Wang2023}, for instance. The interplay between the RKKY and Kondo interactions ultimately yields complex phase diagrams in which both effects can become quantum critical under extrinsic tuning parameters such as pressure, magnetic field or chemical substitution \cite{Coleman2005,Paschen2021}. Thus, accurate classifications of quantum criticality in rare-earth metals often require microscopic probes that can disentangle the contributions arising from long-range or local interactions, respectively.

Inelastic neutron scattering is a momentum-$\bf{Q}$ and energy-$E$ resolved technique with a cross section that is proportional to the imaginary part of the dynamic susceptibility $\chi^{\prime\prime}(\textbf{Q},E)$. This enables precise separation of quantum fluctuations arising from different locations in reciprocal space. This has been particularly important to shed new light on a long-standing conundrum in the community \cite{Poudel_2019,Schroder_2000,Schroder_1998,Stockert_2007,Lohneysen_2007}. Notably, the material CeCu$_{6}$ hosts quantum fluctuations around two different wavevectors $\bf{Q_1}$ and $\bf{Q_2}$ \cite{Amato1988}, and becomes quantum critical under 10\% Au and 20\% Ag substitution on the Cu site \cite{Poudel_2019,Schroder_2000,Schroder_1998,Stockert_2007,Lohneysen_2007}. Pioneering inelastic neutron scattering experiments provided evidence for the existence of anomalous quantum critical fluctuations in CeCu$_{5.9}$Au$_{0.1}$  \cite{Schroder_2000,Schroder_1998}. At larger Au ($>$10\%) and Ag ($>$20\%) concentration, members of the series establish magnetic long-range order that is suppressed in a SDW QPT  under magnetic field \cite{Poudel2015,Stockert_2007,Lohneysen_2007}. A new prospective of this  behavior was provided through a recent study on the zero-filed QPT of CeCu$_{5.8}$Ag$_{0.2}$ that profited from the latest generation of neutron spectrometers \cite{Poudel_2019}. This study investigated a large $(\textbf{Q},E)$ volume with high momentum and energy resolution along with sufficient statistics to distinguish fluctuations from $\bf{Q_1}$ = (0.65, 0, 0.3) and $\bf{Q_2}$ = (1, 0, 0)  in reciprocal lattice units (rlu). The experimental results revealed distinct behavior for the quantum fluctuations at $\bf{Q_1}$ and $\bf{Q_2}$. Notably, the authors noticed that the fluctuations at $\bf{Q_2}$ exhibit a small gap, but that the ones at $\bf{Q_1}$ condense into the elastic line and become quasielastic. The latter fluctuations also revealed a successive increase of the fluctuation life time with decreasing temperature, signifying critical slowing down of the quantum fluctuations which is a fingerprint of divergent imaginary time fluctuations around quantum critical points. Moreover, the dynamic properties at finite temperature provided evidence that the QPT in paramagnetic CeCu$_{5.8}$Ag$_{0.2}$ is also driven by SDW fluctuations \cite{Poudel_2019}.

In this article, we explore these recent insights by addressing the question of how external magnetic field affects the fluctuations in CeCu$_{5.8}$Ag$_{0.2}$ at $\bf{Q_1}$ and $\bf{Q_2}$. This is motivated by earlier studies of CeCu$_6$ showing a pronounced anisotropy in the magnetic susceptibility as a function of the magnetic field direction \cite{Amato1987} and suppression of the magnetic excitations at moderate fields along the crystallographic $c$-axis \cite{Amato1988}. Further macroscopic studies of CeCu$_{6-x}$Au$_{x}$ and CeCu$_{6-x}$Ag$_{x}$ have identified the $c$ axis as the easy axis and the $b$ axis as the hardest axis in the series \cite{Schlager1993, Scheidt2002, Amato1987}. Our microscopic studies of CeCu$_{5.8}$Ag$_{0.2}$ reveal that the quantum critical fluctuations are sensitive to magnetic fields along the  $c$-axis, but remain unchanged at least up to $\mu_0H$ = 8 T for $H||b$. The result reflects the spin anisotropy of the long-range ordered ground state at larger Ag substitution \cite{Poudel2015}, therefore supporting that the quantum critical fluctuations in CeCu$_{5.8}$Ag$_{0.2}$ arise from a SDW QPT.

%----------------------------------------------------------------------
\section{Experimental Details}

Our study was conducted on the same well-characterized CeCu$_{5.8}$Ag$_{0.2}$ single crystal as used in Ref. \cite{Poudel_2019} with $a$ = 8.173, $b$ = 5.098 and $c$ = 10.243 \AA. Two neutron spectrometers were used: (1) LET at  the ISIS Neutron and Muons Source, United Kingdom \cite{Bewley_2011} and (2) CAMEA at the Paul Scherrer Institute (PSI), Switzerland \cite{Camea_2023, Lass_2020} to map large reciprocal space regions for $H||b$ and $c$. The time-of-flight instrument LET was operated with 280 Hz and high-flux slits at the Resolution Chopper and 140 Hz at the Pulse Remover Chopper  to produce neutrons with an incident energy $E_i$ = 3.7 meV,  yielding a resolution of 70 $\mu$eV at the elastic line. The sample was oriented with the $c$-axis vertical and cooled to a temperature $T$ = 250 mK using a dilution insert in a vertical 9 T-cryomagnet. Three-dimensional reciprocal-space maps were recorded with 1 degree sample rotation steps, and chosen such that $\bf{Q_1}$ and $\bf{Q_2}$ were accessible in the large area detector, taking advantage of the out-of-plane detector coverage the instrument provides. The background was measured at a field $\mu_0H$ = 5 T and $T$ = 250 mK. The dynamic susceptibility tensor $\chi^{\prime\prime}(\textbf{Q},E)$ was derived from the measured dynamical structure factor $S(\textbf{Q},E)$ through $\chi^{\prime \prime}(\textbf{Q},E) \propto (1-e^{-\frac{E}{k_BT}}) S(\textbf{Q},E)$ \cite{Furrer_2009}. The temperature dependence ($0.1 \le T \le 4$ K) of the critical fluctuations were measured with the multiplexing spectrometer CAMEA using a $\mu_0H$ = 1.8 T horizontal magnet with large opening windows and a dilution insert. Two dimensional reciprocal maps were measured with 1 degree sample rotation steps and incident energies, $E_i$, of 3.5 and 4.5 meV selected with a double-focusing HOPG monochromator.  This configuration collects excitations up to $E$ $<$ 1.4 meV at an energy resolution of of 110 $\mu$eV at the elastic line. The background was recorded at zero field and $T$ = 45 K. The field dependence along the $b$-axis was studied with CAMEA using a dilution insert in a vertical 11 T-cryomagnet. ($\bf{Q}$, $E$)-maps were acquired at $T$ = 50 mK with the same configuration as described above and fields ranging from $\mu_0H$ = 0 to 8 T. The temperature dependence at $\mu_0H$ = 8 T was measured between $T$ = 0.175 and 3.18 K.

%----------------------------------------------------------------------
\section{Results}

\begin{figure}
\centering
\includegraphics[width=1\linewidth,clip]{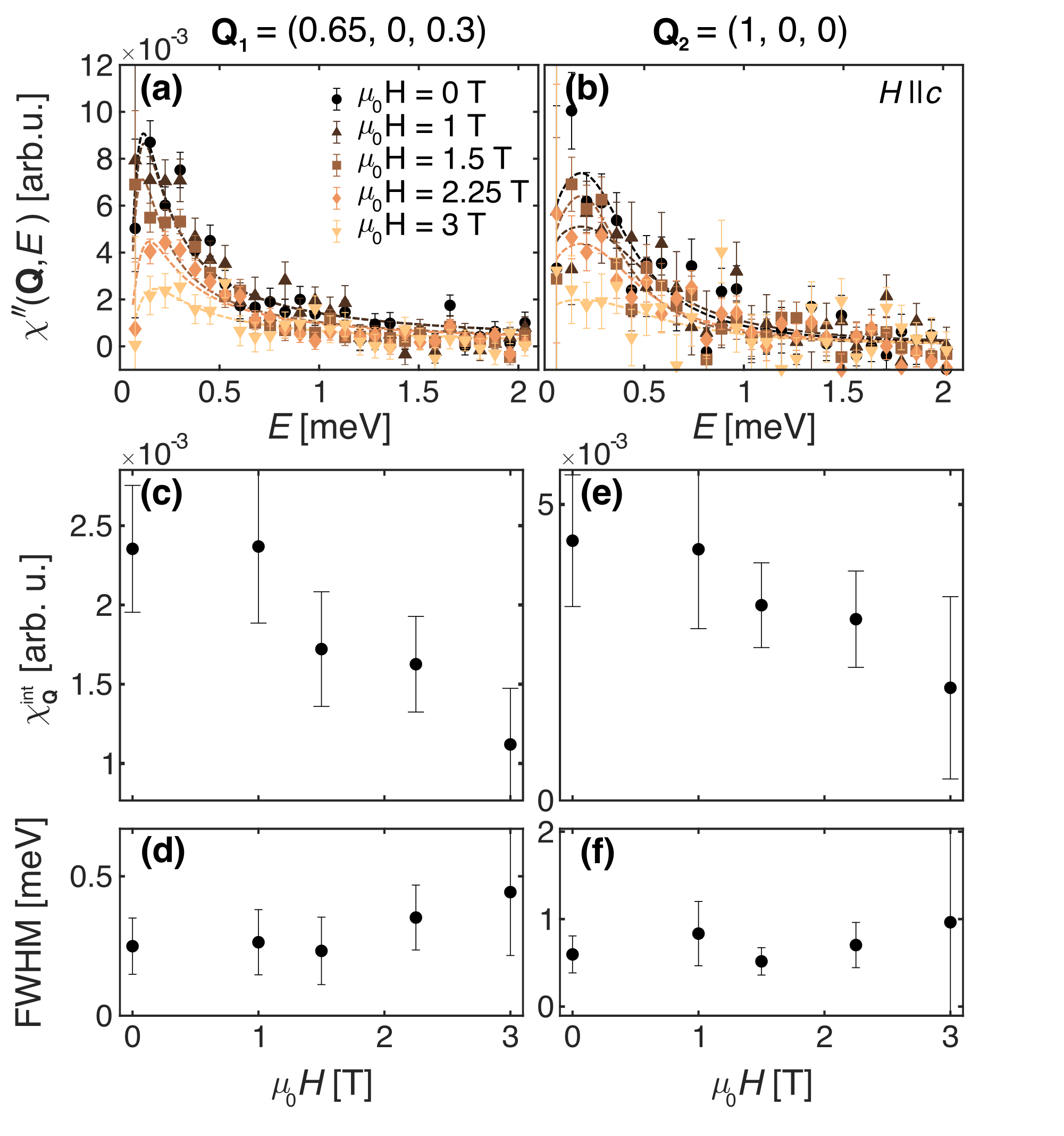}
\caption{Background-subtracted imaginary part of the dynamic susceptibility $\chi^{\prime \prime}(\textbf{Q},E)$ at \textbf{(a)} $\bf{Q_1}$ = ($\pm$0.65, 0, $\pm$0.3) in reciprocal lattice units (rlu) and \textbf{(b)} at $\bf{Q_2}$ = (1, 0, 0) as function of energy transfer $E$. The results were obtained at LET. Each data point was measured at $T$ = 250 mK and integrated over 0.04 rlu$^2$ in the $(H, 0, L)$-plane, 0.4 rlu along the $(0, K, 0)$-axis and over a 0.075 meV energy window. Spectra collected at $\mu_0H$ = 5 T were used as a background (see SM Note 1 for details). \textbf{(c-f)} Integrated intensity $\chi_{\textbf{Q}}^{int}$ and full-width at half-maximum (FWHM) extracted from modified quasielastic-Lorentzian fits at $\bf{Q_1}$ and Lorentzian fits at $\bf{Q_2}$ (see text for details).}
\label{fig:LET}
\end{figure}

The magnetic excitation spectrum measured at LET in zero field and $T$ = 250 mK is shown in the Supplemental Material (SM) Note 1 \cite{Supp_Mat}. Figures \ref{fig:LET}a and b depict the background-subtracted imaginary part of the dynamic susceptibility $\chi^{\prime \prime}(\textbf{Q},E)$ at $\bf{Q_1}$ = ($\pm$0.65, 0, $\pm$0.3) and $\bf{Q_2}$ = (1, 0, 0) as function of energy transfer and for fields applied along the $c$-axis ranging from $\mu_0H$ = 0 to 3 T at $T$ = 250 mK. The $\mu_0H$ = 5 T dataset was used as background, as we found that the fluctuations are suppressed at this field (see SM Note 1). At each \textbf{Q}-position we integrated over 0.04 rlu$^2$ in the $(H, 0, L)$-plane, 0.4 rlu along the $(0, K, 0)$-axis. The intensity at $\bf{Q_1}$ results from an average over the four equivalent $\bf{Q}$-positions to enhance statistics. The energy dependence is modeled through a modified quasielastic Lorentzian
\begin{equation}
    \chi^{\prime\prime}(\textbf{Q},E) = \frac{1}{\sqrt{3}} \frac{(E - E_0)\chi_{\textbf{Q}}^{int}}{(E - E_0)^2 + (\frac{\Gamma}{2})^2}
    \label{equ:quasi_Lorentzian}
\end{equation}
which is normalized such that $\chi^{int}_{\textbf{Q}}$ is the integrated intensity and the full-width at half maximum (FWHM) equals $\sqrt{3}\Gamma$. $E_0$ is an energy offset to test whether the excitation is indeed quasielastic. We obtained $E_0$ = 49(9) $\mu$eV, which is within the elastic resolution of the instrument. The fit is shown as dashed line in Fig. \ref{fig:LET}a, and the resulting fitting parameters are reported in Fig. \ref{fig:LET}c and d. A similar analysis is performed at $\bf{Q_2}$ using a Lorentzian
\begin{equation}
    \chi^{\prime\prime}(\textbf{Q},E) = \frac{\frac{\Gamma}{4}\chi_{\textbf{Q}}^{int}}{(E - E_0)^2 + (\frac{\Gamma}{2})^2}
    \label{equ:Lorentzian}
\end{equation}
which is also normalized such that $\chi^{int}_{\textbf{Q}}$ is the integrated intensity, the FWHM equals $\Gamma$, and $E_0$ = 0.19(5) meV, reporting a gapped excitation. The field dependence of the parameters are reported in Fig.\ref{fig:LET}e and f. We conclude that the quantum fluctuations in CeCu$_{5.8}$Ag$_{0.2}$ rapidly weaken for fields along the $c$-axis and are suppressed at moderate fields of $\mu_0H$ = 5 T.

\begin{figure*}[tbh]
\centering
\includegraphics[width=\linewidth,clip]{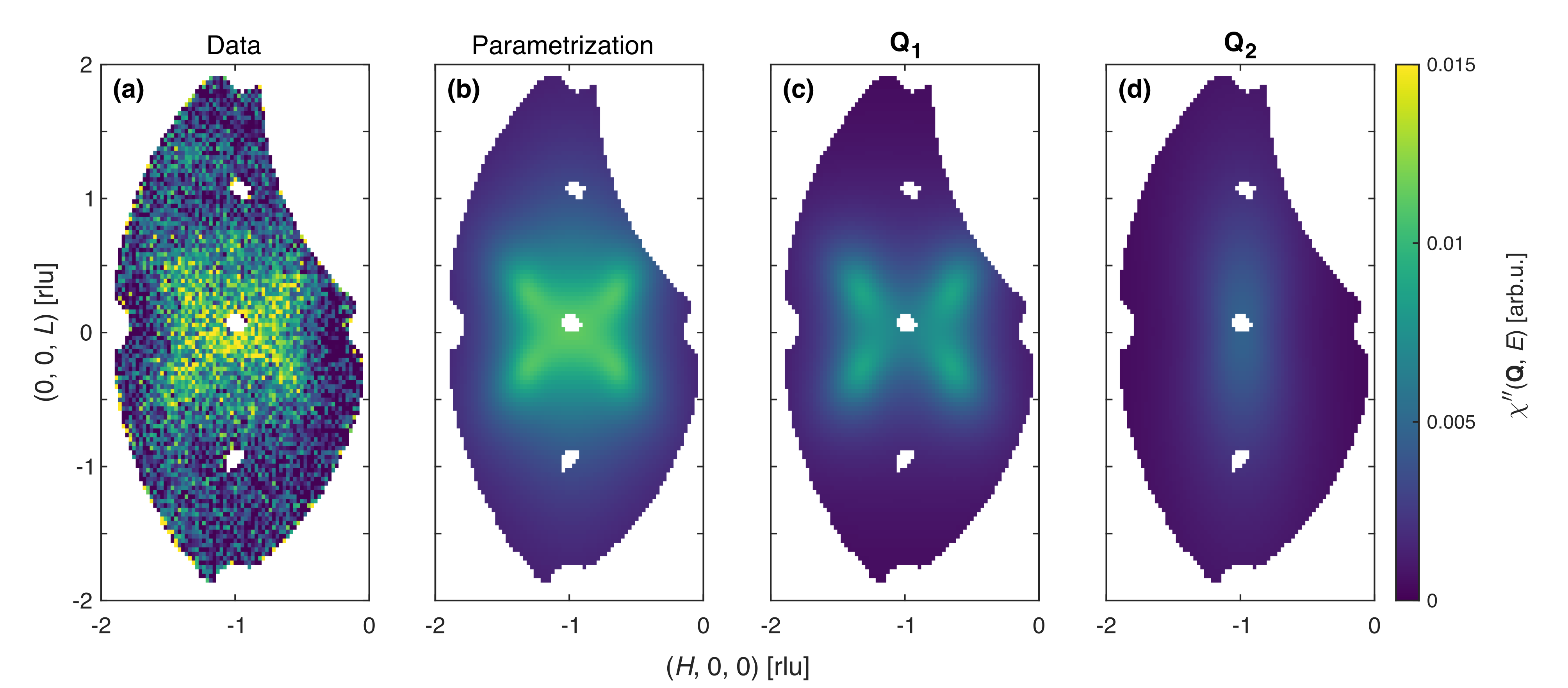}
\caption{\textbf{(a)} Background-subtracted constant energy slice ($E$ integrated from 0.15 to 0.3 meV) in the ($H$, 0, $L$)-plane measured at $T$ = 50 mK with CAMEA. \textbf{(b)}-\textbf{(d)} Parametrization of the experimental data using a combination of two-dimensional Lorentzians for the fluctuations at $\bf{Q_1}$ = ($\pm$0.65, 0, $\pm$0.3) and $\bf{Q_2}$ = (1, 0, 0) (see text for details).}
\label{fig:CAMEA_2Dfit}
\end{figure*}

In all other experiments we collected large $(\textbf{Q},E)$-maps in the ($H$, 0, $L$)-plane at CAMEA
to unambiguously separate the contributions of the fluctuations at $\bf{Q_1}$ and $\bf{Q_2}$. A representative energy slice integrated from $E$ = 0.15 - 0.3 meV is shown in Figure \ref{fig:CAMEA_2Dfit}a. The data is background subtracted and we removed  Bragg-like signatures at $\bf{Q}$ = (-1, 0, $\pm$1) and $\bf{Q_2}$ to highlight the broad fluctuations present in the material. The Figure shows the fluctuations originating at $\bf{Q_1}$ and $\bf{Q_2}$ contribute both to energy cuts at specific $\bf{Q}$-points. Thus, we separated the two signals by the same approach as reported in Ref. \cite{Poudel_2019}. The signal was parametrized with five two-dimensional Lorentzians of the form
\begin{equation}
    y = \frac{A_{\textbf{Q}}(E)}{1 +(\xi_{\parallel}^2\cos^2\theta + \xi_{\perp}^2\sin^2\theta)[(Q_H - Q_{H_i})^2 + (Q_L - Q_{L_i})^2]}
    \label{equ:2D_Lorentzian}
\end{equation}
 where $y$ = $\chi^{\prime\prime}(\textbf{Q},E)$, $A_{\textbf{Q}}(E)$ is the signal amplitude, $\xi_{\parallel}$ and $\xi_{\perp}$ are the correlation lengths of the fluctuations, and $\theta$ is the angle between an arbitrary point $(Q_H, 0, Q_L)$  and $(Q_{H_i}, 0, Q_{L_i})$ which is $\bf{Q_1}$ = ($\pm$0.65, 0, $\pm$0.3) or $\bf{Q_2}$ = (1, 0, 0) (see SM Note 2 for details). Cross correlations effects between different fitting parameters were minimized through an iterative fitting procedure. Notably the reference angle $\theta_0$ required to define the orientation of $\xi_{\parallel}$ and $\xi_{\perp}$ (see SM Note 2) was found using a large energy window. All four correlation lengths were then fixed to their mean values resulting in $\xi_{\parallel}$ = 30(7) \AA\ and $\xi_{\perp}$ = 46(7) \AA\ at $\bf{Q_1}$,  and along the crystallographic $a$- and $c$-axes $\xi_{a}$ = 23(4) \AA\ and $\xi_{c}$ = 12(3) \AA\, respectively, at $\bf{Q_2}$. After this procedure the amplitudes at $\bf{Q_1}$ and $\bf{Q_2}$ were fitted, leading to the corresponding model shown in Figure \ref{fig:CAMEA_2Dfit}b alongside the individual $\bf{Q_1}$ and $\bf{Q_2}$ contributions in Fig. \ref{fig:CAMEA_2Dfit}c and d.

\begin{figure*}[tbh]
\centering
\includegraphics[width=\linewidth,clip]{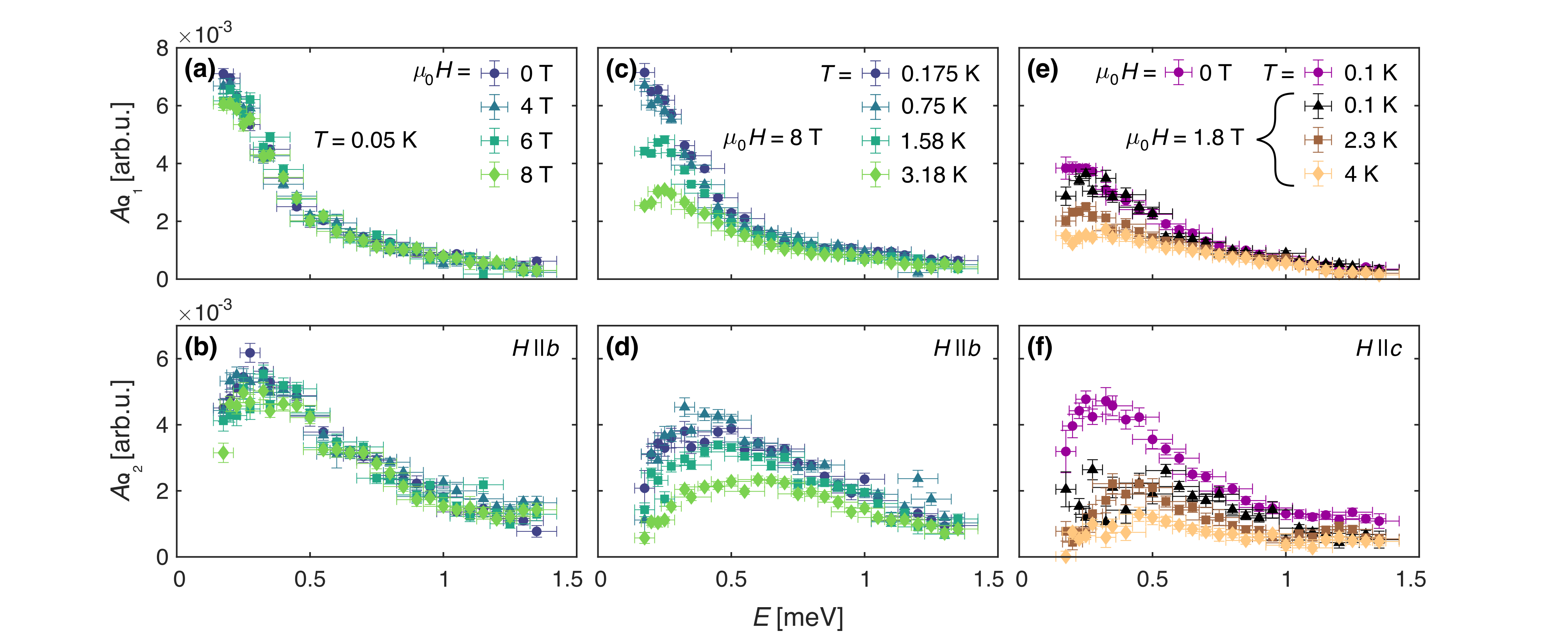}
\caption{Energy dependence of the fluctuation amplitude $A_{\textbf{Q}}$($E$) at $\bf{Q_1}$ = ($\pm$0.65, 0, $\pm$0.3) [panels \textbf{(a)}, \textbf{(c)} and \textbf{(e)}] and $\bf{Q_2}$ = (1, 0, 0) [panels \textbf{(b)}, \textbf{(d)} and \textbf{(f)}] using the parametrization in Eq. \ref{equ:2D_Lorentzian}. In panels \textbf{(a)} and \textbf{(b)} the magnetic field dependence of $A_{\textbf{Q}}$ is shown for a field applied along the crystallographic $b$-axis at $T$ = 50 mK. \textbf{(a)} and \textbf{(b)} report the temperature dependence for a field of $\mu_0H$ = 8 T applied along the $b$-axis. Panels \textbf{(e)} and \textbf{(f)} summarize the field and temperature dependence for a magnetic field along the the $c$-axis.}
\label{fig:CAMEA_2Dfit_results}
\end{figure*}

The field dependence of the fluctuations at $\bf{Q_1}$ and $\bf{Q_2}$ for $H||b$ and $T$ = 50 mK is shown in Fig. \ref{fig:CAMEA_2Dfit_results}a and b, respectively. Consistent with the results reported in Ref. \cite{Poudel_2019}, we observe that the fluctuations at $\bf{Q_1}$ are quasielastic at zero magnetic field whereas at $\bf{Q_2}$ the fluctuations are gapped (see also SM Note 3). However, in contrast to the results obtained for $H||c$ we find that a magnetic field up to $\mu_0H$ = 8 T has no effect on the quantum fluctuations for $H||b$, suggesting a field anisotropy of the quantum fluctuations in the material. This anisotropy of the fluctuations was further studied through their temperature dependence at $\mu_0H$ = 8 T for $H||b$ and $\mu_0H$ = 1.8 T for $H||c$. The corresponding results are shown in Fig. \ref{fig:CAMEA_2Dfit_results}c and d and Fig. \ref{fig:CAMEA_2Dfit_results}e and f, respectively. We find that the fluctuations at $\bf{Q_1}$ increase in intensity and become longer lived as the temperature decreases, suggesting a quantum critical slowing down of the fluctuations (see also SM Note 3). This is not observed for the fluctuations at $\bf{Q_2}$, which increase in intensity and move to lower energy transfers for $H||b$ as the temperature is decreased (further details are reported in SM Note 3). The different behavior of the quantum fluctuations at $\bf{Q_1}$ and $\bf{Q_2}$ as function of temperature provides evidence that the quantum critical fluctuations in CeCu$_{5.8}$Ag$_{0.2}$ originate from fluctuations at $\bf{Q_1}$.

\begin{figure}
\centering
\includegraphics[width=1\linewidth,clip]{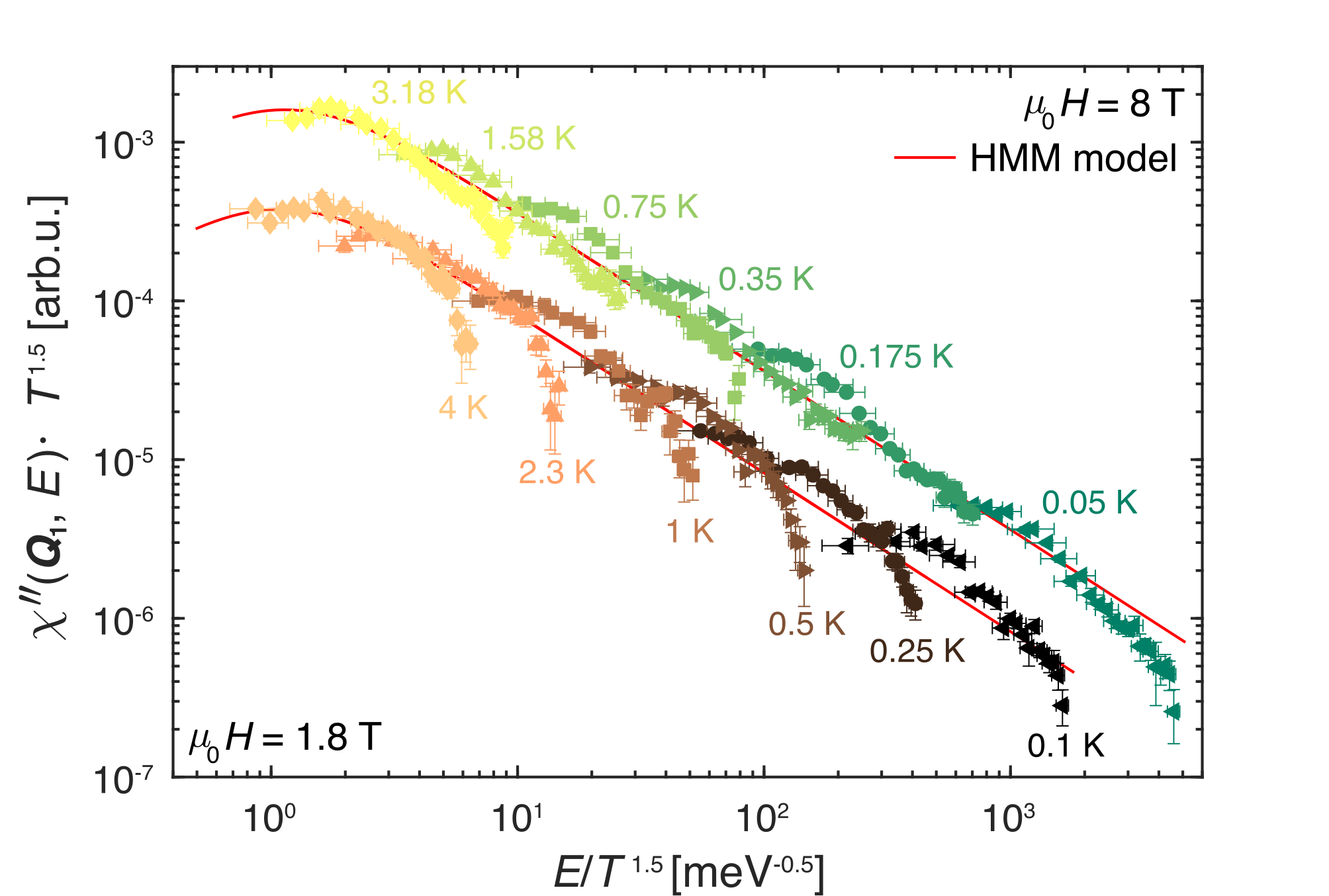}
\caption{Scaling analysis of the fluctuations centered at $\bf{Q_1}$ = ($\pm$0.65, 0, $\pm$0.3) using a HMM model with $\alpha = \beta$ = 1.5. The upper dataset (yellow to green) was measured at $\mu_0H$ = 8 T along the $b$-axis, for the lower dataset (orange to black) $\mu_0H$ = 1.8 T is applied along the $c$-axis (the data points were vertically offset for clarity). The red lines represent the optimal fits to their respective data.}
\label{fig:ET_Scaling}
\end{figure}

The temperature dependence of the critical fluctuations at $\bf{Q_1}$ allows us to assess their nature through an analysis of their scaling behavior as function of $E$/$T$. This is possible because the higher-dimensional fluctuations close to QPTs yield scaling properties $\chi^{\prime\prime}(\textbf{Q}, E) \cdot T^{\alpha}$ =  $f(\textbf{Q},E/T^{\beta})$ of universality classes specific to the nature of the quantum fluctuations. Notably, $\alpha = \beta$ = 1.5 and $f(x) = abx/[1 + (bx)^2]$ is expected for a SDW quantum critical point (HMM model) \cite{Stockert_2007, Schroder_1998,Schroder_2000,Hertz_1976,Millis_1993,Moriya_1995}, but $\alpha$ = 0.72 - 0.83, $\beta$ = 1 and $f(x) = c \sin(\alpha \arctan(x))/[{(x^2 + 1)^{\alpha/2}}]$ for a quantum critical Kondo breakdown (local model) \cite{Stockert_2007,Si_2013,Si_2001, Glossop2007, Grempel2003, Zhu2007}. We present the best fit to our data in Fig. \ref{fig:ET_Scaling}, and report further details in SM Note 4. Our results show that the HMM model ($R^2$ = 0.978 for $H||b$ and $R^2$ = 0.969 for $H||c$) does a better job scaling the data than the local model  ($R^2$ = 0.878 for $H||b$ and $R^2$ = 0.865 for $H||c$), and we refer to the next section for further discussions on the scaling behavior in \CCA{}.

%----------------------------------------------------------------------
\section{Discussion}

The anisotropic field-dependent response of the quantum critical fluctuations provides new information about the nature of the QPT in \CCA{}. The experimental data show that the quantum fluctuations in the material are susceptible to magnetic fields aligned along the $c$-axis, but remain unchanged at least up to $\mu_0H$ = 8 T for $H||b$. The result is in agreement with the magnetic field anisotropy of the magnetic long-range order at larger Ag concentrations and the magnetic susceptibility of the CeCu$_{6-x}$Ag$_{x}$ and CeCu$_{6-x}$Au$_{x}$  series, identifying the orthorhombic $c$- and $b$-direction as the soft and hard axis, respectively \cite{Heuser1997, Schlager1993, Scheidt2002, Amato1987}. This suggests that the quantum critical fluctuations in CeCu$_{6-x}$Ag$_{x}$ with $x$ = 0.2 are spin fluctuations connected to the long-range ordered magnetic moments at $x$ $>$ 0.2, which are amplitude modulated along the $c$-axis \cite{Poudel_2019}. Following this argument we hypothesize that the magnetic exchange couplings in CeCu$_{6-x}$Ag$_{x}$ are anisotropic and much weaker along the $c$-axis when compared to the ones in the $ab$-plane. While this would also be in line with the quasi two-dimensional fluctuations that are thought to be relevant in the CeCu$_6$ family \cite{Kuchler2004,Lohneysen1999}, it remains unclear how that anisotropy arises from the crystal-electric field scheme and the Fermi surface \cite{Witte2007,Goremychkin1993,Chapman_1990,Harima1992}. This encourages reinvestigations of the Fermi surface topology and neutron scattering experiments  clarifying the ground state Hamiltonian of the CeCu$_6$ family, as they may help to further elucidate the connection of the quantum fluctuations with the spin degree of freedom.

Our analysis of the critical scaling behavior for $H||b$ at $\mu_0H$ = 8 T and $H||c$ at $\mu_0H$ = 1.8 T reveals that both datasets scale according to the HMM model with $R^2$ = 0.978 and 0.969, respectively. However, upon visual inspection of the results shown in Fig. \ref{fig:ET_Scaling} a better collapse of the data onto a single curve is observed for $H||b$ when compared to $H||c$. This suggests that the latter dataset is close to a conventional metallic regime in which HMM universality class is no longer valid (see also SM Note 6 for detailed discussion). This statement is supported by macroscopic measurements of family members with larger Ag concentration, showing that regular Fermi liquid behavior is restored above $\mu_0H\gtrapprox$ 2.3 T for CeCu$_{5.2}$Ag$_{0.8}$ and $H||c$ \cite{Heuser1997,Scheidt2002}. This hypothesis can be tested with a study of the scaling behavior for $H||c$ at larger magnetic field strengths. However, technical challenges in neutron instrumentation restricts the feasibility of such experiments and may be achievable only in the future (see SM Note 5 for further information). Additional studies on other CeCu$_{6-x}$Ag$_{x}$ and CeCu$_{6-x}$Au$_{x}$ family members may be also beneficial to clarify nature of the fluctuations at $\bf{Q_2}$ and why they shift with increasing temperature. We suspect that these fluctuations are related to Fermi surface effects.

Finally we argue that a decomposition of competing fluctuations into individual contributions via large ($\bf{Q}$, $E$)-maps can be also advantageous in other cases. Many studies of metals containing rare-earth elements have reported results which are conflicting with prevailing  theories of quantum fluctuations \cite{Park2002,Aronson1995,Stock2012}. Here it is interesting to note that these studies have frequently focused on a few selected points in reciprocal space.  In turn, those materials will benefit from a reinvestigation using state-of-the art neutron spectrometers. Moreover, the method may be also favorable to understand how quantum critical fluctuations are affected by magnetic frustration. Conceptually, it has been predicted that the complex phase diagrams spanned by the RKKY and Kondo interaction can be extended by magnetic frustration \cite{Coleman2010,Vojta_2018,Si_2013}. This has the potential to yield novel QPTs that produce unexplored phenomena such as metallic quantum spin liquid phases \cite{Tokiwa2015,Zhao2019,Tokiwa2013}.

%----------------------------------------------------------------------
\section{Summary}

The magnetic field dependence of the critical fluctuations in CeCeu$_{5.8}$Ag$_{0.2}$ was studied via inelastic neutron scattering. We find that the fluctuations are suppressed rapidly for fields aligned along the orthorhombic $c$-axis, but remain robust up to at least $\mu_0H$ = 8 T for $H||b$. We separated the quantum fluctuations into two contributions at $\bf{Q_1}$ = ($\pm$0.65, 0, $\pm$0.3) and $\bf{Q_2}$ = (1, 0, 0), and observe that only the fluctuations at $\bf{Q_1}$ are quantum critical. Their temperature dependence at $\mu_0H$ = 8 and 1.8 T for $H||b$ and $H||c$, respectively, scales in accordance to the HMM model. Coupled with the observed anisotropy of the quantum critical fluctuations the study suggests a SDW QPT in  CeCeu$_{5.8}$Ag$_{0.2}$, reflecting the spin anisotropy of the long-range ordered state at larger Ag concentrations.

\section*{Acknowledgments}
We acknowledge the Paul Scherrer Institute and the ISIS Neutron and Muon Source for the allocated beam time on CAMEA and LET. We thank the Swiss National Science Foundation for financial support (Grant No. 200021\_200653). The work by ADC (contributions to project leadership, data interpretation, and manuscript preparation), DG Mandrus and FR was supported by the U.S. Department of Energy, Office of Science, Basic Energy Sciences, Materials Sciences and Engineering Division.

The experimental data that support the findings of this article are available under Ref. \cite{data_LET,data_CAMEA}.

% add papers from supplementary that are not cited in the main
\nocite{Knafo2005,Mantid,Horace}

%----------------------------------------------------------------------
%\bibliographystyle{unsrt}
\bibliography{sample}

%------------------------------------------------------------------------------------------------
%------------------------------------------------------------------------------------------------

\newpage

% Reset counter and rename figures
\newcommand{\beginsupplement}{
        \setcounter{table}{0}
        \renewcommand{\thetable}{S\arabic{table}}
        \setcounter{figure}{0}
        \renewcommand{\figurename}{\textbf{Supplemental Figure}}}

\graphicspath{{./}{Supp. Mat./}}
\newcommand{\RM}[1]{\MakeUppercase{\romannumeral #1{}}}

\beginsupplement

\title{Supplemental Material for: Magnetic field dependence of the spin fluctuations in CeCu$_{5.8}$Ag$_{0.2}$}
\date{\today}

\begin{abstract}

\end{abstract}

\pacs{}
%\maketitle\enlargethispage{3pt}
\maketitle{}

\onecolumngrid{

\section*{Supplemental Note 1. LET Experiment Details \& Analysis}

The field dependent critical  fluctuations for $H||c$ were recorded with the time-of-flight instrument LET at the ISIS Neutron and Muons Source, United Kingdom \cite{Bewley_2011}. The crystal was aligned in the horizontal ($H$, $K$, 0)-scattering plane using a copper plate and thin copper wires. We measured with an incoming energy $E_i$ = 3.7 meV providing an elastic line resolution of 70 $\mu$eV. Acquisitions were taken with 1 degree sample rotation steps over 100 degrees around the (1, 0, 0) Bragg position. Each point was measured for a proton current of 5 $\mu$A ($\approx$ 8.5 minutes) at fields $\mu_0H$ = 0, 1, 1.5, 2.25, 3, 5 T. A vertical cryomagnet with dilution insert was used to reach a temperature $T$ = 250 mK. The data reduction was performed with Mantid \cite{Mantid} (version 6.4.0) and the analysis with Horace \cite{Horace} (version 4.0.2).

In Supplementary (Suppl.) Figure \ref{fig:LET_background}a and b, we show the raw data at zero field in the ($H$, $K$, 0)- and  ($H$, 0, $L$)-plane, taking advantage of the out-of-plane coverage of the instrument. The data were integrated over an energy transfer $E$ = 0.15-0.35 meV and $Q_\perp$ = $\pm$0.2 in reciprocal lattice units (rlu) perpendicular to the plane. The data show that while we observe $Q$-dependent fluctuations at $\bf{Q_2}$ = (1, 0, 0) the signal-to-noise ratio did not allow us to unambiguously resolve the  $Q$-dependence of the fluctuations at $\bf{Q_1}$ = ($\pm$0.65, 0, $\pm$0.3). We, thus, performed a box integration centered around $\bf{Q_1}$ and $\bf{Q_2}$ with integration widths  $Q$ = 0.2 rlu in the $HL$-plane and 0.4 rlu along $K$. The signal at $\bf{Q_1}$ was averaged of the four equivalent positions. The results shown in Suppl. Fig. \ref{fig:LET_background}c and d provide evidence that the low-energy fluctuations at $\bf{Q_1}$ and $\bf{Q_2}$ respectively are suppressed at $\mu_0H$ = 5 T.

\begin{figure*}[tbh]
\centering
\includegraphics[width=1\linewidth,clip]{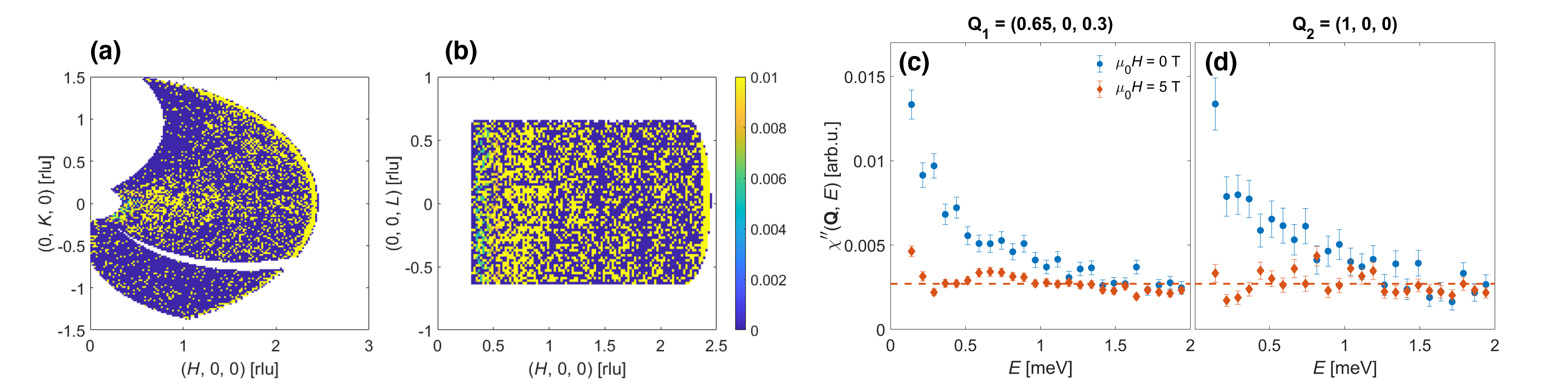}
\caption{Zero-field raw data of the \textbf{(a)} ($H$, $K$, 0) and \textbf{(b)} ($H$, 0, $L$)-plane measured at LET. The data were integrated over an energy transfer window $E$ = 0.15-0.35 meV and out-of-plane momentum transfer $Q_{\perp}$ = $\pm$ 0.2 in reciprocal lattice units (rlu) using an in-plane pixel size $0.02 \times 0.02$ rlu. \textbf{(c)} Fluctuations at $\bf{Q_1}$ = ($\pm$0.65, 0, $\pm$0.3) and \textbf{(d)} $\bf{Q_2}$ = (1, 0, 0) in rlu. Plotted is the imaginary part of the susceptibility $\chi^{\prime \prime}(\textbf{Q},E)$ as function of energy transfer $E$ for $\mu_0$H = 0 and 5 T obtained by box integration around the relevant $Q$-points. The dashed line represents the average value of the $\mu_{0}H$ = 5 T data set.}
\label{fig:LET_background}
\end{figure*}

\section*{Supplemental Note 2. CAMEA Experimental Details \& Parametrization}

The field dependence of the fluctuations for $H||b$ was measured on the multiplexing spectrometer CAMEA at the Paul Scherrer Institute (PSI), Switzerland \cite{Camea_2023,Lass_2020} using a vertical 11 T cryomagnet with dilution insert. The crystal was aligned in the horizontal ($H$, 0, $L$) scattering plane and mounted on a reinforced copper plate using copper wires. The sample was rotated over 115 degrees around (1, 0, 0) in 1 degree steps. Measurements were performed at a temperature $T$ = 0.05 K with incoming energies $E_{i}$ = 3.5 and 4.5 meV at a detector position $2\theta$ = -40 degrees for fields $\mu_{0}H$ = [0, 2, 4, 6, 8] T. Each setup consisted of various scans between which the detector position was shifted by 4 degrees and incoming energies were slightly changed ($\sim$ 0.05 meV) to cover all  blind spots between the analyzers units. At each field, a monitor $M_{3.5}$ = [3'000'000, 3'000'000, 1'500'000, 3'000'000, 1'875'000] and $M_{4.5}$ = [1'840'000, 920'000, 920'000, 920'000, 920'000] for $E_{i}$ = 3.5 and 4.5 meV was used, yielding total counting times $\tau$ = [19, 15, 10, 15, 11] min per point. A separate dataset at $T$ = 45 K was used as background. The temperature dependent data for the scaling analysis were performed on CAMEA for both field directions. The $H||b$ data set was measured with reciprocal maps at $T$ = [0.175, 0.35, 0.75, 1.58, 3.18] K with $\mu_{0}H$ = 8 T. The other instrument parameters were kept identical as the ones described above. The monitor for both incoming energies $M_{3.5}$ = [1'500'00, 1'743'000, 1'743'000, 3'460'000, 3'616'000] and $M_{4.5}$ = [1'000'000, 1'162'000, 1'162'000, 2'308'000, 2'414'000] account for total counting time $\tau$ = [10, 11, 11, 22, 23] min per point. A background was taken at $\mu_{0}H$ = 8 T and $T$ = 45 K. The $H||c$ data set was recorded with  reciprocal maps at $T$ = [0.1, 0.25, 0.5, 1, 2.3, 4] K and $\mu_{0}H$ = 1.8 T using a horizontal 1.8 T cryomagnet with dilution insert. We used the same instrument parameters as above with $M_{3.5}$ = [1'500'000, 1'500'000, 1'500'000, 1'500'000,  1'500'000, 2'353'500] and $M_{4.5}$ = [1'000'000, 1'000'000, 1'000'000, 500'000, 1'000'000, 1'569'000], which amounts to total counting times $\tau$ = [10, 10, 10, 8, 10, 15] min per point. A background was taken at $T$ = 42 K and $\mu_{0}H$ = 1.8 T.

The data reduction for all CAMEA experiments was performed with the MJOLNIR software package \cite{Lass_2020} (version 1.3.1). Bragg contributions were masked with circles of $\Delta Q$ = 0.08-0.15 \AA$^{-1}$ in diameter. The data were parametrized using a two-dimensional Lorentzian centered around  $\bf{Q_2}$ = (1, 0, 0) and four Lorentzians centered around ($\pm$0.65, 0, $\pm$0.3). 
\begin{equation}
    \chi^{\prime\prime}(Q_{H}, Q_{L}, E) = \frac{A_{\textbf{Q}}(E)}{1 +(\xi_{\parallel}^2\cos^2\theta + \xi_{\perp}^2\sin^2\theta)[(Q_H - Q_{H_i})^2 + (Q_L - Q_{L_i})^2]}
    \label{equ:2D_Lorentzian}
\end{equation}
Equation \ref{equ:2D_Lorentzian} describes the  imaginary part of the susceptibility $\chi^{\prime \prime}(\textbf{Q},E)$ at reciprocal position ($Q_H$, 0, $Q_L$) away from a center ($Q_{H_i}$, 0, $Q_{L_i}$) which will be $\bf{Q_1}$ or  $\bf{Q_2}$. $\xi_{\parallel}$ and $\xi_{\perp}$ are the correlation lengths  parallel and perpendicular to the axis intersecting $\bf{Q_1}$ = ($Q_{H_1}$, 0, $Q_{L_1}$) at an angle $\theta_{0}$ from ($H$, 0, 0) (see also Suppl. Fig. \ref{fig:parametrization}). $\theta$ is the angle between the axis along $\xi_{\parallel}$ and  ($Q_H$-$Q_{H_1}$, 0, $Q_L$-$Q_{L_1}$) while $A_\textbf{Q}$ is the amplitude of the signal. Further  information can be found in the Supplementary Information of Ref. \cite{Poudel_2019}.

\begin{figure*}[tbh]
\centering
\includegraphics[width=0.5\linewidth,clip]{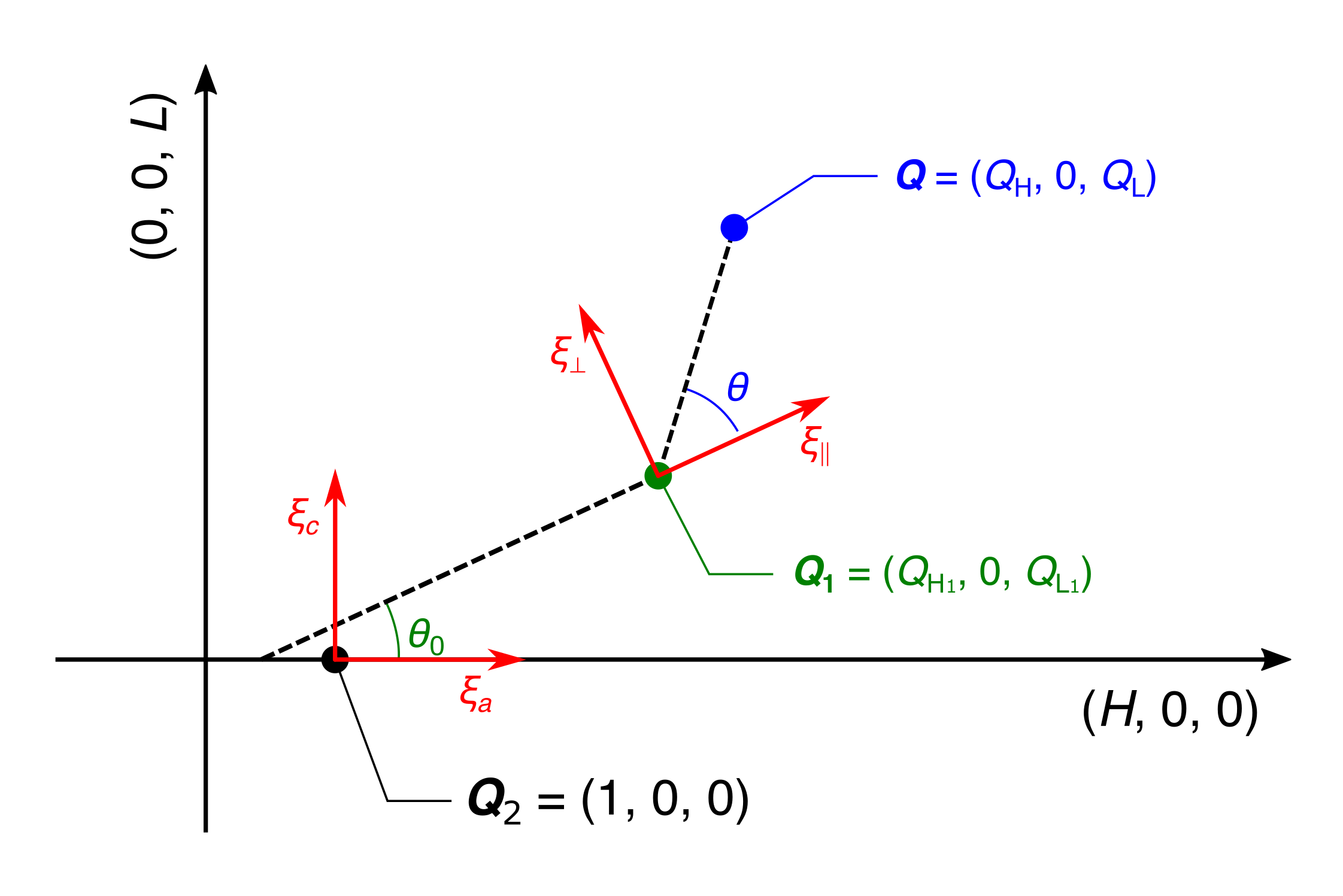}
\caption{Schematic description of the reciprocal space geometry used to parametrize the data at ($Q_H$, 0, $Q_L$) with respect to $\bf{Q_1}$ = ($Q_{H_1}$, 0, $Q_{L_1}$). $\xi_{\parallel}$ and $\xi_{\perp}$ are the correlations lengths parallel and perpendicular to $\bf{Q_1}$ defined by the angle $\theta_0$ with respect to the ($H$, 0, 0)-axis. $\theta$ is the angle representing a counterclockwise rotation  from the axis defined by ($Q_H$-$Q_{H_1}$, 0, $Q_L$-$Q_{L_1}$).}
\label{fig:parametrization}
\end{figure*}

An iterative analysis approach was taken to minimize the effect of cross correlations between free parameters Notably, some of the fitted parameter were fixed in subsequent fitting procedures. A large integration width $E$ = 0.15 - 0.45 meV and square pixel size of side 0.02 \AA$^{-1}$ was used to fix $\theta_{0}$ = 127$^{\circ}$. Smaller energy windows of 0.15 meV were then used to fix the correlations lengths to the mean value of each individual energy cuts. We obtained $\xi_{\parallel}$ = 30(7) \AA\ and $\xi_{\perp}$ = 46(7) \AA\ at $\bf{Q_1}$,  and along the crystallographic $a$- and $c$-axes $\xi_{a}$ = 23(4) \AA\ and $\xi_{c}$ = 12(3) \AA\, respectively, at $\bf{Q_2}$. During the last step of the fitting procedure, only the amplitude of both fluctuations were left as free parameters. The result of this fit is shown in Fig. 2 of the main document.

\section*{Supplemental Note 3. Detailed Temperature and Field Dependence}

The results shown in Figure 3 of the main document were further analyzed by fitting the data points to a modified quasielastic Lorentzian (Equation \ref{equ:quasi_Lorentzian}) and Lorentzian (Equation \ref{equ:Lorentzian}) for the fluctuations at  $\bf{Q_1}$ and $\bf{Q_2}$, respectively. The functions are defined such that $\chi_{\textbf{Q}}^{int}$ is the integrated intensity and  the Full-Width at Half Maximum (FWHM) given by $\sqrt{3}\Gamma$ and $\Gamma$, respectively. $E_{0}$ serves as an energy offset.

\begin{equation}
    \chi^{\prime\prime}(\textbf{Q},E) = \frac{1}{\sqrt{3}} \frac{(E - E_0)\chi_{\textbf{Q}}^{int}}{(E - E_0)^2 + (\frac{\Gamma}{2})^2}
    \label{equ:quasi_Lorentzian}
\end{equation}

\begin{equation}
    \chi^{\prime\prime}(\textbf{Q},E) = \frac{\frac{\Gamma}{4}\chi_{\textbf{Q}}^{int}}{(E - E_0)^2 + (\frac{\Gamma}{2})^2}
    \label{equ:Lorentzian}
\end{equation}

The results are shown in Suppl. Fig. \ref{fig:CAMEA_quasielastic/lorentzian} where $\chi_{\textbf{Q}}^{int}$, FWHM and energy offset $E_{0}$ are plotted as function of either field or temperature for magnetic field directions along $b$ and $c$. In fits without offsets, $E_{0}$ featured no clear dependence and was fixed to the mean value $E_0$ = 0.107(6) meV for the temperature dependence with field along the $c$-axis, 0.128(3) meV for the temperature dependence with field along the $b$-axis and 0.132(4) and 0.23(2) meV for the field dependence of the $\textbf{Q}_1$ and $\textbf{Q}_2$ fluctuations, respectively. These values were fixed to improve the fit of  $\chi_{\textbf{Q}}^{int}$ and FWHM. The green line corresponds to the FWHM of a Bragg peak while the black line is a linear fit of $E_0$. We note that the energy offset $E_0$ at $\textbf{Q}_1$ is within or close to the elastic resolution of the CAMEA spectrometer (110 $\mu$eV), supporting the interpretation that these fluctuations are quasielastic. The fluctuations at $\textbf{Q}_1$ show a
pronounced temperature dependence of the fluctuation lifetime (see FWHM of the fluctuations in
Suppl. Fig. \ref{fig:CAMEA_quasielastic/lorentzian}b and k), suggesting a quantum critical slowing down of the fluctuations in
paramagnetic CeCu$_{5.8}$Ag$_{0.2}$. This provides evidence that the quantum critical fluctuations in the material originate at $\textbf{Q}_1$. We mention that at very low temperature critical slowing down of quantum fluctuations around quantum critical points is often difficult to
observe in neutron scattering. This is because it becomes increasingly more challenging to separate the
fluctuations from the elastic line as the Lorentzian width decreases with decreasing temperature.
Thus, often the width of the quantum fluctuations appears to not completely vanish, and efforts in
precisely modeling the instrument resolution function and measurements with better energy
resolution are required to observe the critical slowing down at very low temperatures (see for
instance Ref. \cite{Stockert_2007}). Such a detailed analysis has not been done in our case, because of potential systematic errors in modeling the resolution function within the 2d analysis we employ here. Furthermore, it is well established that CeCu$_{5.8}$Ag$_{0.2}$ at zero magnetic field features quantum critical fluctuations, and therefore features critical slowing down of the fluctuations \cite{Lohneysen_2007}. Thus, the saturating FWHM at low temperature arises either from systematic errors of the fit close to the elastic line or indicates deviations from the quantum critical point as magnetic field is applied (see also Suppl. Note 6 discussing limitations of our analysis) \cite{Knafo2005}. However, the field dependence of the quantum fluctuations for $H||b$ shown in Suppl. Fig. \ref{fig:CAMEA_quasielastic/lorentzian}f and g reveal no observable deviations from the zero-field behavior, supporting an interpretation that the saturation of the FWHM at low temperature arises from systematic errors rather than intrinsic effects.

\begin{figure*}[tbh]
\centering
\includegraphics[width=1\linewidth,clip]{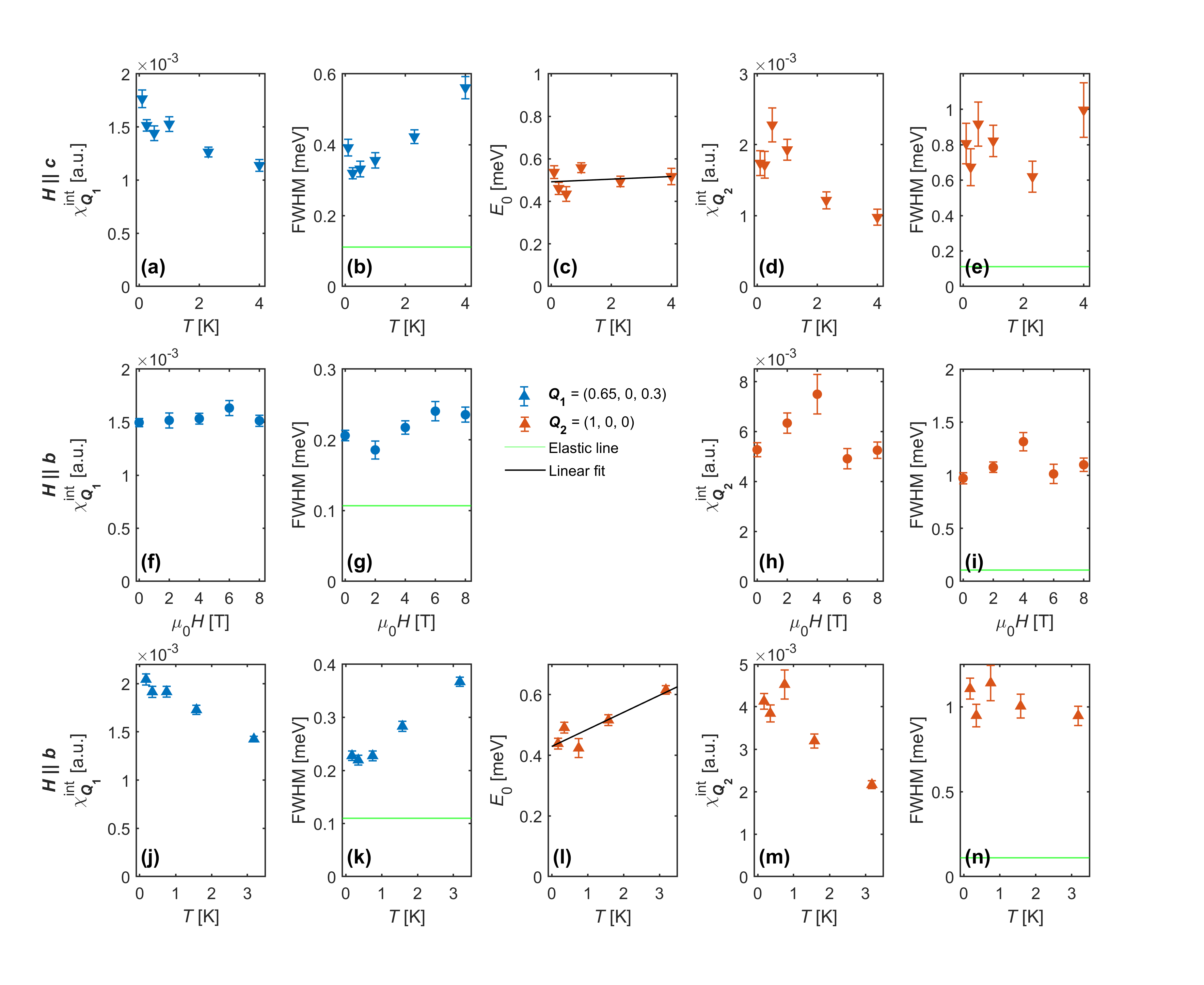}
\caption{\textbf{(a)} and \textbf{(b)} Integrated intensity $\chi_{\textbf{Q}}^{int}$ and FWHM as function of temperature extracted from a quasielastic Lorentzian fit at $\bf{Q_1}$ for a field of $\mu_0H$ = 1.8 T along $c$. \textbf{(c)}-\textbf{(e)} Energy offset $E_0$, integrated intensity $\chi_{\textbf{Q}}^{int}$ and FWHM extracted from a Lorentzian fit at $\bf{Q_2}$. The green line represents the resolution of the instrument and the black line is a linear fit of the offset. \textbf{(f)}-\textbf{(i)} Same as \textbf{(a)}-\textbf{(e)}  for field along $b$ at $T$ = 50 mK. \textbf{(j)}-\textbf{(n)} Same as \textbf{(a)}-\textbf{(e)} for temperature dependent data for a field of $\mu_0H$ = 8 T along $b$.}
\label{fig:CAMEA_quasielastic/lorentzian}
\end{figure*}

\newpage
\section*{Supplemental Note 4. Critical Scaling Analysis}

In this section we compare the goodness of fit between the Hertz, Millis, Moriya (HMM) model and the local model.  In Suppl. Figure \ref{fig:critical_exponent}a, the goodness of fit is reported as function of critical exponents $\alpha$ and $\beta$, such that $\chi^{\prime\prime}(\textbf{Q},E) \cdot T^{\alpha} = f(E/T^{\beta})$. The function $f$ is given by Eqs. \ref{equ:f_HMM} and \ref{equ:f_local} for the HMM and Local model respectively \cite{Stockert_2007, Schroder_1998,Schroder_2000,Hertz_1976,Millis_1993,Moriya_1995,Si_2013,Si_2001}. Critical exponents $\alpha = \beta = 1.5$ are expected for the HMM model, while the local model is expected to lie in a broader region around $0.72 < \alpha < 0.83$ and $\beta = 1$ \cite{Glossop2007, Grempel2003, Zhu2007}.

\begin{equation}
    f(x) = \frac{abx}{1 + (bx)^2}
    \label{equ:f_HMM}
\end{equation}

\begin{equation}
    f(x) = \frac{c \sin(\alpha \arctan(x))}{(x^2 + 1)^{\alpha/2}}
    \label{equ:f_local}
\end{equation}

The optimized R-square for both models and field directions is represented by the white dot. While they are not in the immediate proximity to the expected values, the R-square is very close to the expected value for the HMM model, indicating good agreement with the model. In contrast, the expected exponents of the local model have a substantially worse R-square than the extracted optimal value. In comparison, the HMM scenario yields an R-square of 0.978, and 0.878 for the local scenario $\alpha$ = 0.83 and field along $b$. The data measured with field along $c$ yields similar values of 0.969 and 0.865 for the HMM and local model, respectively. In Suppl. Figure \ref{fig:critical_exponent}b we show the scaling behavior of the data using the local model.

\begin{figure*}[tbh]
\centering
\includegraphics[width=0.85\linewidth,clip]{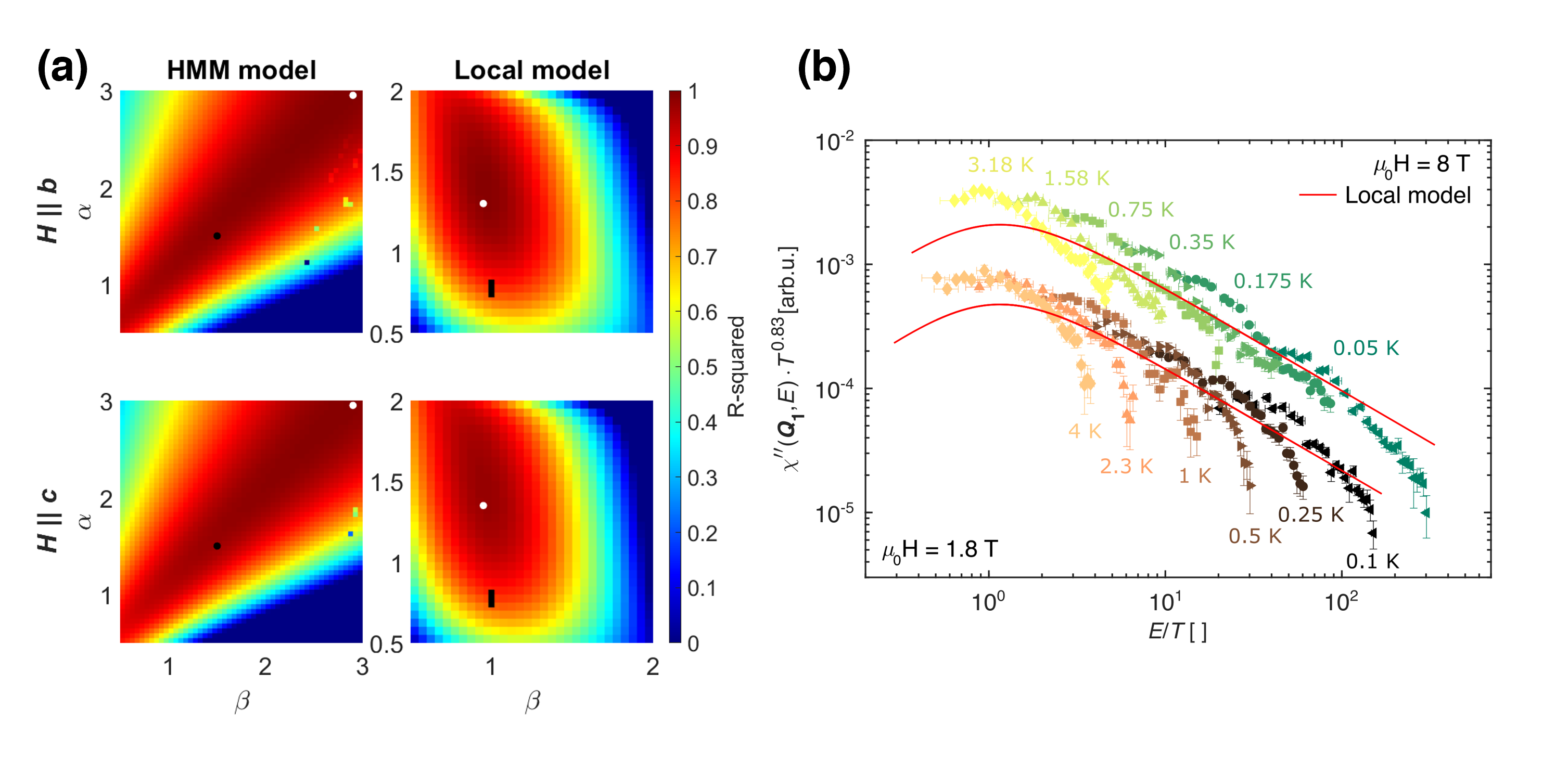}
\caption{\textbf{(a)} R-square value of the ($E$/$T$) scaling fit using the HMM and local model for field along $b$ (top row) and field along $c$ (bottom row) as function of $\alpha$ and $\beta$. The white dot represent the best R-square. The black dot and line represent the values of $\alpha$ and $\beta$ expected for the two models, $i.e.$ $\alpha = \beta = 1.5$ for the HMM model, and $\beta = 1$ and $0.72 < \alpha < 0.83$ for the local model. \textbf{(b)} Scaling analysis of the fluctuations centered at $\bf{Q_1}$ = ($\pm$0.65, 0, $\pm$0.3) using a local model with $\alpha$ = 0.83, $\beta$ = 1. The upper dataset (yellow to green) was measured at $\mu_0H$ = 8 T along the $b$-axis, for the lower dataset (orange to black) $\mu_0H$ = 1.8 T is applied along the $c$-axis (the data points were vertically offset for clarity). The red lines are optimized fits to their respective data.}
\label{fig:critical_exponent}
\end{figure*}

\section*{Supplemental Note 5. Current limitations in neutron spectroscopy}

The data shown in Fig. 4 of the main text suggest that the fluctuations measured at $H||c$ with $\mu_0H$ = 1.8 T  are close to a conventional metallic regime in which HMM universality class is no longer valid. This hypothesis can be tested with a study of the scaling behavior for $H||c$ at larger magnetic field strengths. However, an unambiguous decomposition for the fluctuations at $\bf{Q_1}$ and $\bf{Q_2}$ require either cold neutron time-of-flight spectrometers with exceptional signal-to-noise ratio to benefit from the out-of-plane coverage and vertical field cryomagnets, or cold neutron multiplexing spectrometers using strong horizontal magnets with large opening windows. The combination of these requirements strongly restricts the feasibility of such experiments and may be achievable only in the future.

\section*{Supplemental Note 6. Limitations in our analysis}

The fluctuations at $\textbf{Q}_1$ show a
pronounced temperature dependence of the fluctuation lifetime (see FWHM of the fluctuations in
Suppl. Fig. \ref{fig:CAMEA_quasielastic/lorentzian}b and k), suggesting a quantum critical slowing down of the fluctuations in
paramagnetic CeCu$_{5.8}$Ag$_{0.2}$. However, the FWHM seems to saturate at temperatures below $T\approx$ 1 K suggesting only a partial slowing down of the quantum fluctuations. We mention that critical slowing down of quantum fluctuations around quantum critical points is often difficult to
observe at low temperature using neutron scattering. This is because it becomes increasingly more challenging to separate the
fluctuations from the elastic line as the Lorentzian width decreases with decreasing temperature.
Thus, often the width of the quantum fluctuations appears to not completely vanish, and efforts in
precisely modeling the instrument resolution function and measurements with better energy
resolution are required to observe the critical slowing down at very low temperatures (see for
instance Ref. \cite{Stockert_2007}). Such a detailed analysis has not been done in our case, because of potential systematic errors in modeling the resolution function within the 2d analysis we employ here. Furthermore, it is well established that CeCu$_{5.8}$Ag$_{0.2}$ at zero magnetic field features quantum critical fluctuations, and therefore features critical slowing down of the fluctuations \cite{Lohneysen_2007}. Thus, the saturating FWHM at low temperature likely arises from systematic errors of the fit close to the elastic line. This is also supported by the field dependence of the quantum fluctuations for $H||b$ shown in Suppl. Fig. \ref{fig:CAMEA_quasielastic/lorentzian}f and g revealing no observable deviations from the zero-field behavior.

A partial slowing down of the quantum fluctuations at low temperature may also indicate that the quantum fluctuations in CeCu$_{5.8}$Ag$_{0.2}$ are no longer quantum critical upon application of magnetic field. We argue that in such a case the scaling analysis would still report on the proximity of the fluctuations to the established scaling behavior at the zero field quantum critical point, which is what we indicate in the discussion stating that the fluctuations for $H||c$ are close to the Fermi liquid regime at $\mu_0H$ = 1.8 T. However, a previous study on quantum critical fluctuations has revealed that in such a case a geometrical construction of the scaling behavior can lead to a false scaling behavior \cite{Knafo2005}. The paper underlines that firstly, only the data points which show critical slowing down should be used (for instance above $T_N$ if the system were ordered), and secondly that false scaling can be observed if $\beta$ is fixed while $\alpha$ is a free parameter in the scaling analysis. While, we varied $\alpha$ and $\beta$ in the scaling analysis (see SM Note 4), the first point is a potential concern if the saturating FWHM at low temperature stems from intrinsic rather than instrument effects. To further clarify potential limitations of the scaling analysis we considered in Suppl. Fig. \ref{fig:CAMEA_quasielastic/Knafoscaling} only experimental data points $T \geq$ 1 K (the value the FWHM appears to saturate in Suppl. Fig. \ref{fig:CAMEA_quasielastic/lorentzian}b and k). We mention that exceeding difficulties to separate the contributions of the two fluctuations close to the background level makes such efforts impossible for temperatures above $T$ $>$ 4 K in reasonable counting times at current neutron facilities. However, despite the limited amount of data, the results support the conclusion that the scaling behavior is in line with an HMM model and that the data at $\mu_0H$ =  8 T for $H||b$ are closer to the quantum criticality than for $\mu_0H$ = 1.8 T at $H||b$ (see $R^2$ values).

\begin{figure*}[tbh]
\centering
\includegraphics[width=0.85\linewidth,clip]{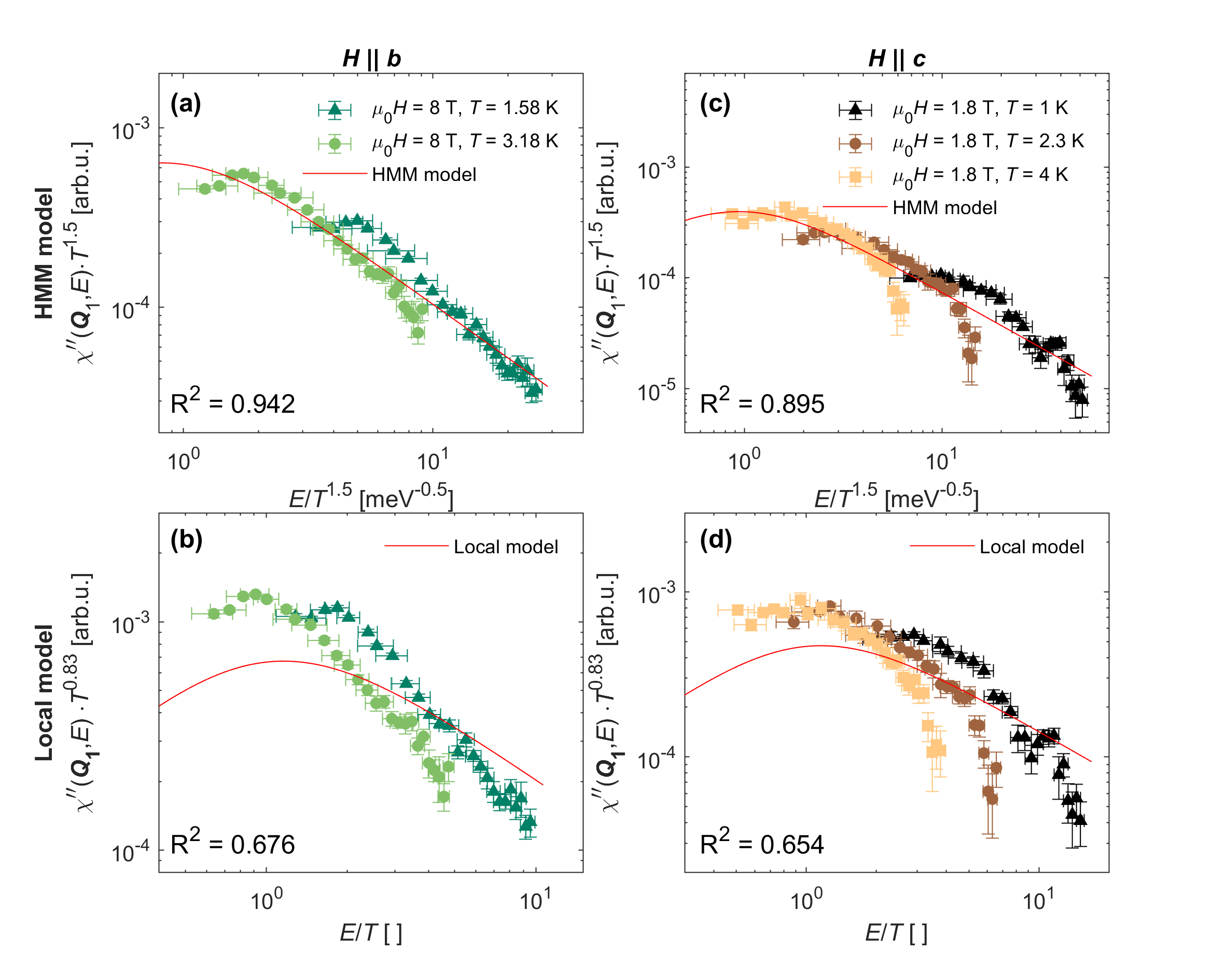}
\caption{Scaling analysis of the fluctuations centered at $\bf{Q_1}$ = ($\pm$0.65, 0, $\pm$0.3) using only the data points $T$ $>$ 1 K using the HMM model with $\alpha$ = $\beta$ = 1.5 for $H||b$ in \textbf{(a)}, the local model with $\alpha$ = 0.83, $\beta$ = 1 for $H||b$ in \textbf{(b)}, the HMM model with $\alpha$ = $\beta$ = 1.5 for $H||c$ in \textbf{(c)} and the local model with $\alpha$ = 0.83, $\beta$ = 1 for $H||c$ in \textbf{(d)}. The red lines are optimized fits using Eq. \ref{equ:f_HMM} and Eq. \ref{equ:f_local}. $R^2$ is the R-square value}
\label{fig:CAMEA_quasielastic/Knafoscaling}
\end{figure*}

}
\end{document}